\documentclass{elsart}

\def\alt{\stackrel{<}{{}_\sim}}
\def\agt{\stackrel{>}{{}_\sim}}

\begin{document}
\input epsf
%\begin{pf*}{ams}
\runauthor{V.A. Kuzmin, I.I. Tkachev}
\begin{frontmatter}
\vskip 0.5in
\hfill  {\em {In honour of Lev Okun, for his continuous inspiration}}\\

\title{Ultra High Energy Cosmic Rays and Inflation Relics.}

\author{Vadim A. Kuzmin$^a$ and Igor I. Tkachev$^{a,b}$}
\address{
$^a$Institute for Nuclear Research, Russian Academy of Sciences,\\
60th October Anniversary Prosp. 7a, Moscow 117312, Russia
and\\
$^b$TH Division, CERN, CH-1211 Geneva 23, Switzerland\\
}
\maketitle

\begin{abstract}
There are two processes of matter creation after inflation that 
may be relevant to the resolution of
the puzzle of cosmic rays observed with energies beyond GZK cut-off: 
1) gravitational creation
of superheavy (quasi)stable particles, and 2) non-thermal phase transitions
leading to formation of topological defects.
We review both possibilities.
\end{abstract}

%\pacs{PACS: 98.70.Sa, 95.35.+d, 98.80.Cq}


\end{frontmatter}

\section{Introduction}

Cosmological and astrophysical considerations are able to provide
the strongest restrictions on parameters of particle physics models
and even rule out some classes of models entirely. This is
especially valuable when the model is unrestricted by laboratory experiments
(which  is often the case). Among famous results which made a strong
impact on model building is the 
cosmological domain wall problem which appears in models with
spontaneous breaking of discrete symmetries \cite{zko} and the problem of
magnetic monopoles in Grand Unified theories \cite{zk}.
In return, studies of cosmological phase transitions \cite{KL}
and of the dynamics of bubbles of a metastable vacuum \cite{KOV} 
lead to the change of
basic concepts of the cosmology of the early Universe, 
and inflationary cosmology
\cite{Strb,guth,al82,al83} was born (for reviews see \cite{al90,KT_book}). 
Inflation gives a possible solution to horizon, flatness and 
homogeneity problems of ``classical'' cosmology \cite{guth}.
Inflation was designed to solve the problem of unwanted relics, like magnetic
monopoles. It was promptly realized \cite{structure} 
that inflation can generate small 
amplitude large scale density fluctuations which are the necessary seeds 
for the galaxy and the large scale structure formation in the Universe. 
This elevates inflation
from the rank of a ``broad brush problem solver'' into the rank of a testable 
hypotheses. And testable in fine details, as rapidly accumulating data
on cosmic microwave background fluctuations (starting from COBE detection
\cite{COBE} through numerous balloon and ground based CMBR 
experiments and with culmination 
at MAP \cite{MAP} and PLANK \cite{PLANK} anticipated detailed maps of 
anisotropy of the microwave sky) and huge 
galaxy catalogs like the already collecting data SLOAN digital sky surview 
\cite{SLOAN} will provide a wealth of cosmological information.

Inflation is generally assumed to be driven by the special
scalar field $\phi$ known as the {\it inflaton}. 
During inflation, the inflaton field
slowly rolls down towards the minimum of its potential. Inflation
ends when the potential energy associated with the inflaton field 
becomes smaller than the kinetic energy, which happens when magnitude
of the inflaton field decreases below the Plank scale, 
$\phi \alt M_{\rm Pl}$ and ``cold'' coherent oscillations of the 
inflaton field commence. These oscillations contained all the energy of 
the Universe at that time. 

All matter in the Universe was created in reheating, which is nothing
but decay of the zero momentum mode of inflation oscillations.  
The process is obviously of such vital importance that here too one 
may hope to find some observable consequences, specific
for the process itself and for particular models of particle physics,
despite the fact that scales relevant for the reheating are
very small.
And, indeed, we now believe that there can be some clues left.
Among those are: topological defects production in non-thermal phase
transitions \cite{nth}, GUT scale baryogenesis \cite{bau},
generation of primordial background of stochastic gravitational waves
at high frequencies \cite{gw}, just to mention a few. However, the
most interesting can be a possible relation
to a mounting puzzle of the Ultra High Energy Cosmic Rays (UHECR) 
\cite{KT98}.

When a proton (or neutron) propagates in CMB, it gradually looses energy 
colliding with photons and creating pions. There is a threshold
energy for the process, so it is effective for very energetic nucleons
only, which leads to the  Greisen-Zatsepin-Kuzmin (GZK) 
cutoff \cite{gzk} of the high energy tail of the
spectrum of cosmic rays. All this means that detection of, say,
$3\times 10^{20}$ eV proton would require its source to be within
$\sim 50$ Mpc. However, several well established events above the cut-off 
were observed
by Yakutsk \cite{Yak}, Haverah Park \cite{HP}, Fly Eye \cite{FE} and 
AGASA  \cite{AGASA1} collaborations
(for the recent reviews see Refs. \cite{BBDGP,bs}).

Results from the AGASA experiment \cite{AGASA} are shown in Fig. 1.
The dashed curve represents the expected  spectrum if conventional 
extragalactic sources of UHECR would be distributed uniformly in the 
Universe. This curve displays the theoretical GZK cut-off, but one 
observes events
which are way above it. (Numbers attached to the data points show the 
number of events observed in each energy bin.) Note that no candidate 
astrophysical source, like powerful active galaxy nuclei, were found in 
the directions of all six events with $E > 10^{20}$ eV \cite{AGASA}
(at these energies cosmic rays experience little deflection by 
galactic magnetic fields).

\begin{figure}
\centerline{\leavevmode\epsfysize=7.cm \epsfbox{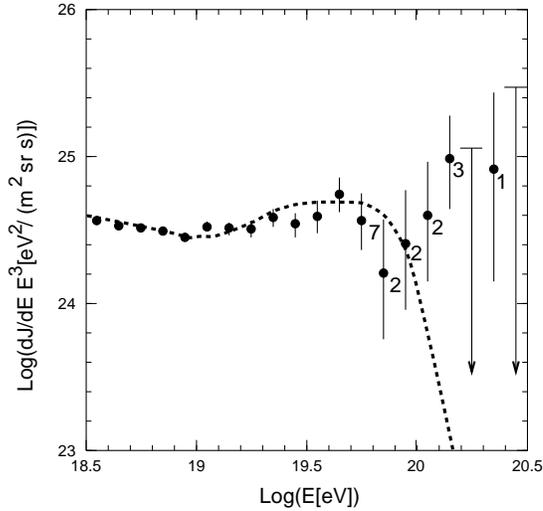}}
\vspace{10pt}
\caption{AGASA data set \cite{AGASA}, February 1990 -- October 1997.}
\label{fig1}
\end{figure}

Is some unexpected astrophysics at work here or this is at last 
an indication of the long awaited new physics ?

There are two logical possibilities to produce UHE cosmic rays:
either charged particles have to be accelerated to energies
$E > 10^{20}$ eV, or UHECR originate in decays
of heavy X-particles, $m_X > 10^{12}$ GeV.
Maximum energy which can be achieved in an accelerating site of the size $R$
which has the magnetic field strength $B$ is \cite{hillas}:
\begin{equation}
E < 10^{20} Z \, \frac{B}{\mu {\rm G}}\, \frac{R}{\rm Mpc}\; {\rm eV} \, \, .
\label{hillas}
\end{equation}
A magnetic field is required either to keep the particle confined within 
the accelerating region or to produce an accelerating electric field.
For protons (Z=1) a few sources satisfy this condition: pulsars, 
active galactic nuclei (AGNs)
and radio-galaxies. However, energy losses (pair production and
meson photoproduction) restrict the maxim energy
to $E < 10^{16}$ eV in pulsars and AGNs \cite{BBDGP,AGNlimit},
while radio-galaxies that lie along the arrival directions of UHECR are 
situated at large cosmological distances, $\agt 100$ Mpc \cite{radio-gal},
i.e. beyond the GZK radius. Similar conclusion seem to be true with respect to 
cosmological Gamma Ray Bursts as a possible source of UHECRs \cite{GRB}.

New astrophysics which may work is a possibility to generate
UHECR within GZK sphere in remnants of dead quasars \cite{boldt}
(these are dormant galaxies which harbour supermassive spinning black hole).

New physics suggested as an explanation of UHE cosmic rays, range up to the
violation of the Lorentz invariance \cite{lorentz}. Among less radical
extensions of the standard model are:
\begin{itemize}
\item  The existence of a particle which is immune to CMB
in comparison with nucleons. 
In this scenario the primary particle is produced in remote 
astrophysical accelerators (e.g. radio-galaxies) and is able to travel 
larger cosmological distances while having energies above the GZK cut-off.  
There are variations to this scheme. 
\begin{itemize}
\item Suprsymmetric partner of gluon, the gluino, can form bound states 
with quarks and gluons. If gluino is light and quasistable 
(see e.g. \cite{farrar,raby}), the
lightest gluino containing baryons will have sufficiently large
GZK threshold to be such a messenger \cite{farrar} and as a hadron it
will be able to produce normal air showers in the Earth's atmosphere.
However, there are strong arguments due to Voloshin and Okun \cite{VO86}, 
against light quasistable gluino based on constraints on the abundance of 
anomalous heavy isotopes which also will be formed as bound states with 
gluino.
\item High energy (anti)neutrinos produced in distant astrophysical
sources will annihilate via $Z^{0}$ resonance on the relic neutrinos 
and produce energetic gamma or nucleon \cite{nu}.
The relic neutrino masses in the eV range are consistent with this
scenario~\cite{GK99}, as well as with the Super-Kamiokande results.
The required high density of the relic neutrinos is achieved if
gravitational clumping takes place~\cite{nu} or if the Universe has a
significant lepton asymmetry in background neutrinos~\cite{GK99}. Even then 
the total luminosity of the neutrino sources in the Universe must be as
high as $10^{-2} - 10^{1}$ of its photon luminosity, and, therefore,
neutrino-only sources are called for by the upper bound from the
flux of the cosmic rays~\cite{BW99}. An independent constraint on the
density of the relic neutrinos comes from
CMBR and already the present data start to be challenging
for models with large neutrino asymmetry~\cite{KR99}.
\end{itemize}
\item Another class of suggestions is related to topological defects. 
UHECR are produced when topological
defects decompose to constituent fields (X-particles) which in turn decay 
\cite{hsw}. Maximum energy is not a problem here, but in models which
involve string \cite{S}  or 
superconducting string \cite{hsw} networks, 
the typical separation between defects is of order of the Hubble distance
and thus these models are subject to GZK cut-off. Models in which defects
can decay ``locally'' include
networks of monopoles connected by strings (necklaces) \cite{necl}, 
vortons (charge and current carrying loops of superconducting strings
stabilized by angular momentum) \cite{vortons}, and
monopolonium (bound monopole-antimonopole pairs) \cite{monium}.
Finally, magnetic monopoles accelerated by intergalactic magnetic fields 
were also considered as primary UHECR particls \cite{mon}. 
\item Conceptually the simplest possibility is that UHECR are produced
cosmologically locally in decays of some new particle \cite{KR,BKV}.
GZK cut-off is automatically avoided but
the candidate $X$-particle must
obviously  obey constraints on mass, number density and lifetime.
\end{itemize}

\section{UHECR from decaying particles}

In order to produce cosmic rays in the energy range
$E > 10^{11}$ GeV, the decaying primary
particle has to be {heavy}, with the mass well above GZK cut-off,
$m_X > 10^{12}$ GeV. The lifetime, $\tau_{X}$, cannot be much smaller than
the age of the Universe, $t_U \approx 10^{10}$~yr. Given this shortest
possible lifetime, the observed flux of UHE cosmic rays will be
generated with the rather low number density of $X$-particles,
$\Omega_{X} \sim 10^{-12}$,
where $\Omega_{X} \equiv m_{X} n_{X}/\rho_{\rm crit}$, $n_X$ is the
number density
of X-particles and $\rho_{\rm crit}$ is the critical density.
On the other hand, X-particles must not overclose the Universe,
$\Omega_{X} < 1$.
With $\Omega_{X} \sim 1$, the X-particles may play the role of cold dark
matter and the observed flux of UHE
cosmic rays can be matched if $\tau_{X} \sim 10^{22}$~yr.

\begin{figure}
\centerline{\leavevmode\epsfysize=7.cm \epsfbox{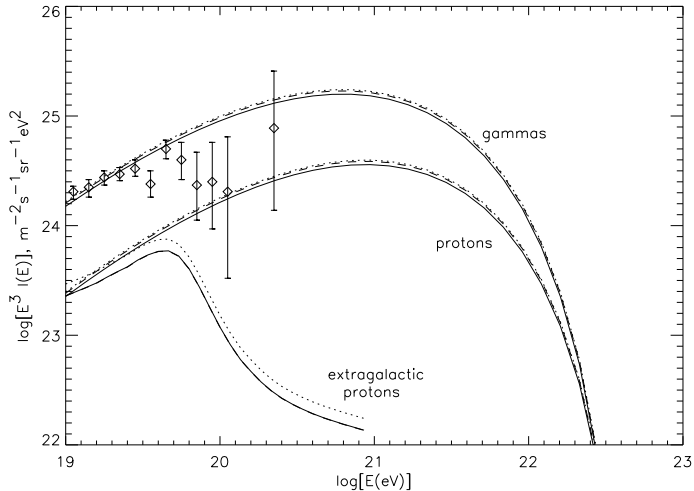}}
\vspace{10pt}
\caption{Predicted fluxes from decaying X-particles, as calculated
in Ref. \cite{WB}  and the data. Latest AGASA results, 
Fig. 1, are not shown.}
\label{berez}
\end{figure}

Spectra of UHE cosmic rays arising in decays of relic X-particles
were successfully fitted to the data 
for $m_X$ in the range $10^{12} < m_X/{\rm GeV} < 10^{14}$
\cite{WB,BS}.
For example, the fit of Berezinskii et. al. \cite{WB} to observed 
fluxes of UHECR assuming
$m_X \approx  10^{14}$ GeV is shown in Fig.~\ref{berez}.  Beside the mass
of the X-particle there is another
parameter which controls the flux of the cosmic rays from decaying particles:
namely, the ratio of X-particles number density and their lifetime. 
For the fit in Fig.~\ref{berez} it was used
$({\Omega_X}/\Omega_{\rm CDM})(t_U/\tau_X) = 5\times 10^{-11}$.

The problem of the particle physics mechanism responsible for a long but
finite lifetime of very heavy particles can be solved in several ways.
For example, some otherwise conserved quantum number carried by  X-particles
may be broken very weakly
due to instanton transitions \cite{KR}, or quantum gravity (wormhole)
effects \cite{BKV}. If instantons are responsible for $X$-particle
decays, the lifetime is estimated as
$\tau_{X} \sim m_{X}^{-1}\cdot \mbox{exp}(4\pi/{\alpha_{X}})$,
where $\alpha_{X}$ is the coupling constant of the relevant gauge
interaction.
The lifetime will fit the allowed window if the coupling constant
(at the scale $m_{X}$) is $\alpha_{X} \approx 0.1$ \cite{KR}.

A class of natural candidates for superheavy long-living particles
which arise in string and M theory was re-evaluated recently in Refs.
\cite{BEN98} and particles with desired mass and long life-time were
identified. Other interesting candidates were found among adjoint
messengers in gauge mediated supergravity models \cite{HYZ98}
and in models of superheavy dark matter with discrete gauge symmetries
\cite{hyy}. Superheavy dark matter candidates in superstrings and supregarvity 
models were considered also in Refs. \cite{othe_models}.

Below we address the issue of X-particle abundance.

\section{Superheavy particle genesis in the early Universe}

Seperheavy particles can be created in the early Universe by several 
mechanisms. Among those are:
\begin{itemize}
\item Non-equilibrium ``thermal'' production in scattering or 
decay processes in primordial plasma \cite{KR,BKV}.
\item Production during decay of inflaton oscillations (``preheating'')
\cite{KLS94,KT97a,KT97b,PR}.
\item Direct gravitational production from vacuum fluctuations during 
inflation \cite{CKR,KT98,KT99}.
\end{itemize}

In any case the final ratio of the density in $X$ particles to the entropy
density is normalized by the reheating temperature. The reheating
temperature is limited to the value below $10^8$ - $10^9$ GeV in supergravity
models with decaying heavy gravitino \cite{gtino}. This restricts model 
parameters when ``thermal'' mechanism of heavy particle production is 
operative (but does not rule it out \cite{KR,BKV,thermal-wimpZ}).

The last two mechanisms are closely related to each other and both can be 
described on equal footing within frameworks of a single unified approach:
particle creation in external time varying background. However, while the 
outcome of the second mechanism is highly dependent upon 
the strength of the interaction
of the X-field to the inflaton, no coupling (e.g. to the inflaton or plasma) 
is needed in the third mechanism when the temporal change of the metric 
is the single cause of particle production. Even absolutely sterile
particles are produced by the third mechanism which may be relevant
for very long-living superheavy particles. Resulting abundance is quite 
independent of the detailed nature of the particle which makes the 
superheavy (quasi)stable X-particle a very interesting dark matter 
candidate.

We concentrate here on the second and third mechanisms and from the start 
we introduce coupling of the X-field to the inflaton for uniformity 
of discussion. The limit of zero coupling will correspond to pure 
gravitational production. 

In the case of a heavy scalar field $X$ we consider the model
\begin{equation}
L={1\over 2} (\partial_{\mu}\phi)^2 - V(\phi )
+{1\over 2} (\partial_{\mu}X)^2 - {1\over 2} (M_X^2 - \xi R) X^2
- {g^{2}\over 2} \phi^{2} X^{2} \; .
\label{sXmodel}
\end{equation}
Here $V(\phi )$ is the inflaton potential. In simple ``chaotic'' \cite{al83}
models of inflation the inflaton is either ``massive'' with the scalar 
potential $V(\phi) = m_\phi^2\phi^2/2$, or ``massless'',
$V(\phi) = \lambda\phi^4/4$. Normalization to the large scale
structure requires $m_\phi^2/M_{\rm Pl}^2 \approx 10^{-12}$ in the former model
and $\lambda \approx 10^{-13}$ in the latter model.
The constant $\xi$ describes direct coupling to the space-time 
curvature $R$, with $\xi=0$ corresponding to the minimal coupling
and $\xi =1/6$ being the case of conformal coupling.

Fermion field (spin $\half$) is conformally coupled to gravity. In addition
to standard kinetic and mass terms it also may have coupling to 
the inflaton,
$V_\half = g \phi \bar{X}X$. 

It is convenient
to work in conformal metric $ds^2 = a(\eta )^2 (d\eta^2 - d{\bf x}^2)$
with rescaled fields, $\varphi \equiv \phi a(\eta)$ and
$\chi \equiv X a(\eta)^s$, where $s=1$ and $s=\threehalf$ 
for scalar and fermion
fields respectively. In what follows we 
measure time and space intervals in units of inflaton mass, 
$\tau \equiv m\eta$.

\subsection{Quantum fields in classical backgrounds}
\label{quantum_fields}

Here we summarise the basic formalism of 
particle creation in external classical background (e.g. space-time
metric of an expanding universe or oscillating inflaton field).
For more details see e.g. Refs. \cite{pc_pwl,MMF,pc,gmm}.

i) {\it Spin 0 bosons.}

A real scalar field is Fourier expanded in a comoving box
\begin{equation}
\chi(\tau, {\bf x}) = \sum_{\bf k} [\chi_{k}(\tau) a_{\bf k} + 
\chi_{k}^*(\tau) a_{\bf -k}^\dagger] e^{ i{\bf kx}}\,\, .
\label{fexpansion}
\end{equation}
Annihilation and creation operators commute except for 
$[a_{\bf k},a_{\bf k}^\dagger] = 1$.
The  mode functions, $\chi_{k} \equiv \chi_{k}(\tau )$
of a scalar Bose field are solutions of the oscillator equation
\begin{equation}
{\chi ''}_{k} + \omega^2_k(\tau) \chi_{k} = 0 \,\, ,
\label{lin}
\end{equation}
with the time dependent frequency
\begin{equation}
\omega^2_k(\tau) = k^2 + \frac{a''}{a} (6\xi-1) + m_{\chi}^2 a^2
+4q\varphi^2  \,\, ,
\label{ome}
\end{equation}
where $' \equiv d/d\tau$ and 
\begin{equation}
m_{\chi}^2 \equiv \frac{M_X^2}{m^2} \,\, , {\hspace{1cm}}
q \equiv \frac{g^2 \phi^2(0)}{4 m^2} \,\, .
\label{mchi}
\end{equation}
Here 
$\phi(0)$ is the value of the inflaton field when it starts to 
oscillate (which corresponds to normalization $\varphi (0) = 1$).

Let $\omega_k$ at some time interval satisfy the adiabatic
condition $|\omega_k'|/\omega_k^2 \ll 1$. We can choose solutions
of Eq. (\ref{lin}) which enter decomposition Eq. (\ref{fexpansion}) 
to be positive-frequency modes 
$\chi_{k} = \omega^{-1/2}_k e^{ -i\omega_k \tau}$ 
and define vacuum state $a_{\bf k}|0\rangle = 0$. The number of particles
in a non-vacuum state will be constant during evolution through this
adiabatic interval. Let the adiabatic condition be violated for some time
and then the system enters another adiabatic interval. In that interval
we can define another set of positive-frequency modes and corresponding
vacuum. Initial positive-frequency modes evolved through non-adiabatic 
region will not coincide with ``out'' state modes, but one set of modes
can be expressed in terms of the other. This decomposition is called
Bogolyubov transformation \cite{Bog_tr}. Since one and the same field
is expanded with the use of two different sets of mode functions,
the Fourier coefficients are also related to each other
\begin{equation}
a_{\bf k}^{\rm out} = 
\alpha_{\bf k} a_{\bf k} + \beta^*_{\bf k} a_{\bf k}^\dagger
\, \, .
\label{BT}
\end{equation}
It follows immediately that the initial vacuum state at late
times contains particles  
\begin{equation}
\langle 0 |a_{\bf k}^{\rm \dagger~out} a_{\bf k}^{\rm out} |0\rangle
= |\beta_k|^2
\, \, .
\label{BT2}
\end{equation}

Technically it is easier to find Bogolyubov's coefficients by diagonalizing
Hamiltonian of the field $X$. For 
any time moment $\tau$ this procedure gives
\begin{equation}
|\beta_k|^2 = \frac{ |{\chi '}_{k}|^2 + \omega^2 |\chi_{k}|^2
-2 \omega } {4\omega} \, ,
\label{beta}
\end{equation}
where mode functions are solutions of Eq. (\ref{lin}) with
initial (vacuum) conditions
\begin{equation}
{\chi }_{k}(0) = \omega^{-1/2},\; \; \;
{\chi '}_{k}(0)=-i\omega \chi_{k} \, .
\label{incon}
\end{equation}

ii) {\it Spin 1/2 fermions.}

The relevant mode functions of the Fermi field satisfy the oscillator
equation with the complex frequency
\begin{equation}
{\chi ''}_{k} + (\omega^2_k - i m_{\rm eff}')\chi_{k}= 0 \,\, ,
\label{lin_f}
\end{equation}
where the real part of the frequency is given by
$\omega^2_k = k^2 + m_{\rm eff}^2$ and 
$m_{\rm eff} = m_\chi a + \sqrt{q} \varphi$.
We choose
\begin{equation}
{\chi }_{k} (0) = \sqrt{1 - \frac{m_{\rm eff}}{\omega} },\; \; \;
{\chi '}_{k} (0) =-i\omega \chi_{k} \, ,
\label{inconf}
\end{equation}
as the initial conditions. In this case we find per spin state
\begin{equation}
|\beta_k|^2 =
\frac{\omega - m_{\rm eff} - {\rm Im}(\chi_{k} \chi^{*'}_{k})}{2\omega} \, .
\label{beta_f}
\end{equation}

Finally, the number density of $X$-particles created by time varying
background is
\begin{equation}
n_X=\frac{1}{2\pi^2a^3} \sum_s\int |\beta_k|^2 k^2 dk \, \, ,
\label{nX}
\end{equation}
where $\sum_s$ is the sum over spin states. 
The expression (\ref{nX}) gives the number
density of particles only, with an equal amount of antiparticles being
created in the case of charged fields. 

\subsection{Gravitational creation of particles}

It was noticed \cite{CKR,KT98} that superheavy particles are produced
gravitationally in the early Universe from vacuum fluctuations and their
abundance can be correct naturally, if the standard Friedmann epoch in
the Universe evolution was preceded by the inflationary stage.
This is a fundamental process of particle creation unavoidable in the
time varying background 
and it requires no interactions. Temporal change of the metric is the single
cause of particle production.
Basically, it is the same process which during inflation had generated
primordial large scale density perturbations. 
No coupling (e.g. to the inflaton or plasma) is needed.
All one needs are
stable (very long-living) X-particles with mass of order of
the inflaton mass, $m_X \approx 10^{13}$ GeV.
Inflationary stage is not required to produce superheavy particles
from the vacuum.
Rather, the inflation provides a cut off in excessive gravitational 
production of heavy
particles which would happen in the Friedmann Universe if it would
start from the initial singularity \cite{KT98}.
Resulting abundance is quite independent of the detailed
nature of the particle which makes the superheavy (quasi)stable X-particle
a very interesting dark matter candidate. 
So, we start our consideration with gravitational creation of particles, 
i.e. we put $g=0$ (or equivalently $q=0$) in the formulas above.

\subsubsection{Friedmann Cosmology }

\begin{figure}
\centerline{\leavevmode\epsfysize=7.cm \epsfbox{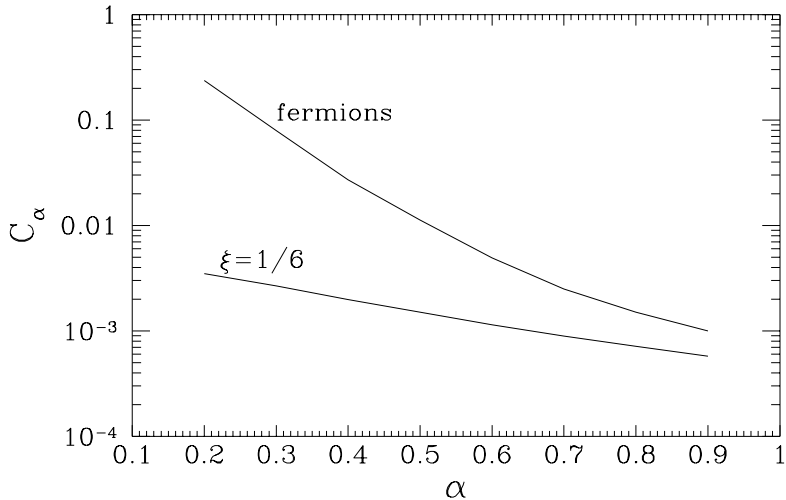}}
\vspace{10pt}
\caption{ The coefficient $C_\alpha$, defined in Eq. (\ref{nx}),
is shown as a function of $\alpha$ for the background cosmology
with a power law scale factor $a \propto t^\alpha$, Ref. \cite{KT99}.
}
\label{sc_alpha}
\end{figure}

For particles with conformal coupling to gravity (fermions
or scalars with $\xi = 1/6$), it is the particle mass 
which couples the system to the
background expansion and serves as the source of particle
creation. Therefore, just on dimensional grounds, we expect 
\begin{eqnarray}
n_X \propto m_X^3 a^{-3} 
\end{eqnarray} 
at late times when particle creation diminishes.
In Friedmann cosmology, $a \propto (mt)^\alpha \propto (m/H)^\alpha$
($\alpha=\half$ and $\alpha=\threehalf$ for radiation and matter dominated
expansion respectively). We conclude that
the anticipated formulae for the X-particles abundance can be 
parameterised as
\begin{eqnarray}
n_X = C_\alpha m_X^3 \left(\frac{H}{m_X} \right)^{3\alpha} \, .
\label{nx}
\end{eqnarray} 
On the other hand, it is expansion of the Universe which 
is responsible for particle creation. Therefore, this equation
which describes simple dilution of already created particles,
is valid when expansion becomes negligible, $H \ll m_X$. This means also
that particles with $m_X \gg H$ cannot be created by this
mechanism. Creation occurs when $H \sim m_X$. The coefficient $C_\alpha$ 
depends upon the background cosmology only, and it 
can be found numerically \cite{KT98}, see Fig.~\ref{sc_alpha}.

In particular, for the radiation dominated Universe,
which was studied also in Ref. \cite{pc_pwl},  one finds 
$\Omega_X \equiv \rho_X/\rho_c = m_X n_X \, 32\pi Gt^2/3$ 
with the present value of $\Omega_X$ being equal to
$\Omega_X \sim 2\times 10^{-2} (m_X^2/M^2_{\rm Pl})\sqrt{m_X t_e}$,
where $t_e$ is the time of equal densities of radiation
and matter in the $\Omega =1$ Universe. This gives
$\Omega_X \sim (m_X/10^9 {\rm GeV})^{5/2}$. Stable
weakly interacting particles with $m_X \agt 10^9$ GeV
will overclose the Universe even if initially they were in a vacuum state
and were created from the vacuum during
the regular radiation dominated stage of the Universe evolution.

There is no room for superheavy particles in our Universe if it
started from the initial Friedmann singularity \cite{KT98},
since the value of the Hubble constant is limited from above only
by the Planck constant in this case.

\subsubsection{Inflationary Cosmology }

However, this restriction will not be valid if inflation separates
initial conditions, whatever they were, from the observable Universe.
In inflationary cosmology the Hubble constant (in effect) did not exceeded
the inflaton mass, $H < m_\phi$. The mass
of the inflaton field has to be $m_\phi \sim 10^{13}$ GeV as constrained
by the amplitude of primordial density fluctuations relevant for 
the large scale structure formation. Therefore, direct gravitational 
production of particles with {$m_X > H \sim 10^{13}$ GeV} has to be 
suppressed in inflationary cosmology.

Particle creation from vacuum fluctuations during inflation
(or in the de Sitter space) was extensively studied \cite{pc_dS1,pc_dS2}, 
usually in the case of small $m_X$ and in application to  generation of 
density fluctuations necessary for the large scale structure formation. The
characteristic quantity which is usually cited in this applications, 
the variance of the field, $\langle X^2 \rangle$, is defined by an 
expression similar to Eq. (\ref{nX}), in the typical case 
$\alpha_k \approx - \beta_k$ the integrand
is being multiplied by the factor $2\sin^2(\omega_k\tau)/\omega_k$.
For example, for the scalar Bose field with the minimal coupling to the
curvature,
$\langle X^2 \rangle = 3H_i^4/8\pi^2m_X^2$ if
$m_X \ll H_i$\cite{pc_dS1,pc_dS2}.  For massless self-interacting field
$\langle X^2 \rangle \approx 0.132 H_i^2/\sqrt{\lambda}$ \cite{SY}.
Particle creation for the specific case of the Hubble dependent 
effective mass, $m_X(t) \propto H(t)$, was considered in Ref. \cite{LR}.

Results  \cite{KT99} of direct numerical integration of 
gravitational creation of superheavy particles
in chaotic inflation model with the potential 
$V(\phi) = m^2_\phi \phi^2/2$ is shown in Fig. 2.

\begin{figure}
\centerline{\leavevmode\epsfysize=7.cm \epsfbox{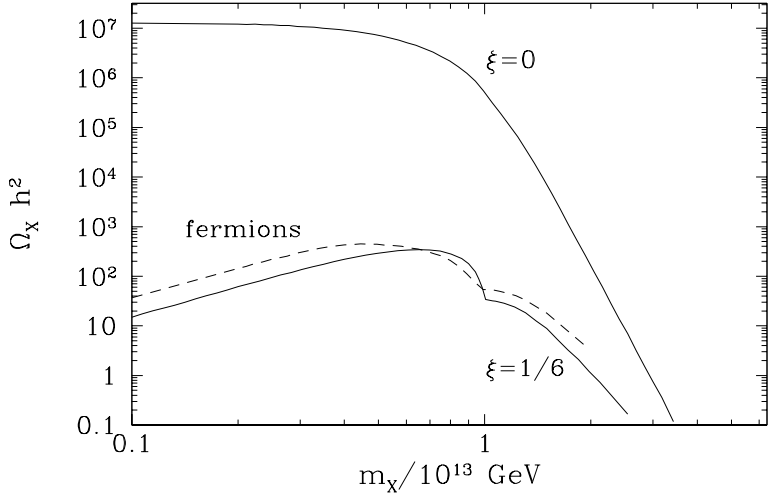}}
\vspace{10pt}
\caption{Ratio of the energy density in $X$-particles, gravitationally 
generated in inflationary cosmology, to the critical energy density is 
shown as a function of X-particle mass, Ref. \cite{KT99}.}
\label{fig:m_infl}
\end{figure}

This figure was calculated assuming $T_{\rm R} = 10^9$ GeV for the reheating
temperature. (At reheating the entropy of the Universe was created in 
addition to 
X-particles. In general, multiply this figure by the ratio
$T_{\rm R}/10^9$ GeV and devide it by the fractional entropy increase 
per comoving volume if it was significant at some late epoch.)
The reheating temperature is constrained, $T_{\rm R} < 10^{9}$ GeV, 
in supergravity theory \cite{gtino}. 
We find that $\Omega_X h^2 < 1$ if
$m_X \approx ({\rm few})\times 10^{13}$ GeV.
This value of mass is in the range suitable
for the explanation of UHECR events \cite{KT98}. 
Gravitationally created superheavy 
X-particles can even be the dominating form of matter in the Universe today 
if X-particles are in this mass range \cite{CKR,KT98}. 

\subsubsection{Isocurvature fluctuations in superheavy particle matter}

In numerical calculations, Ref. \cite{KT99}, it was found
that the variance $\langle X^2 \rangle$
of the field $X$  measured at the end of inflation is independent upon
$m_X$ if the mass of $X$ is small and coupling to curvature is minimal.
At some later epoch when $H \approx m_X$ (which will be long after
the end of inflation if X is a light field)
the field $X$ starts to oscillate on all scales, including $k=0$.
Only at this time, which we denote by  $t_X$, all field fluctuations are
transformed into the non-zero particle density and we can use
$\rho_X = m_X n_X \approx m_X^2 \langle X^2 \rangle $.
The variance of $X$ fluctuations
was unchanged on large scales, starting from the end of inflation
down to the
time $t_X$. So, when the field starts to oscillate
$\rho_X \propto m_X^2$. However, the  energy density of the inflaton
field,
$\rho = 3H^2/8\pi G$, decreased during this time interval in proportion
to $H^2(t_X)/H^2(0) \approx m^2_X/H^2(0)$. That is why the ratio of
the energy
density in $X$-particles to the total energy density does not depend
on $m_X$ when measured at $t > t_X$, see Fig. \ref{fig:m_infl}.

\begin{figure}
\centerline{\leavevmode\epsfysize=7cm \epsfbox{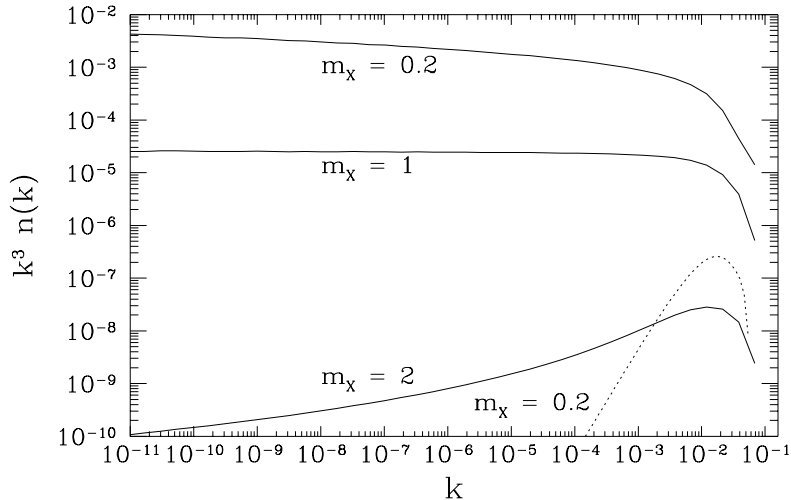}}
\caption{Spectrum of created particles, $k^3n(k)$, in a model with massive
inflaton is shown for several choices of the mass of scalar
X-particle with
the minimal coupling (solid lines) and the conformal coupling
(dotted line), Ref. \cite{KT99}.
Masses and momenta, $k$, are given in units of the inflaton mass.
}
\label{fig:nk}
\end{figure}

Variance of the  field X is different from the usually
calculated for the fixed de Sitter inflationary background because
we consider the actual evolution of the scale factor and the value of
the Hubble parameter is not constant
during inflation, being larger at earlier times. Correspondingly,
the number
of created particles per decade of $k$ grows logarithmically towards
small k if $m_X$ is small. (Power spectrum behaves similarly.)
The  examples of the particle number,
$k^3 n_X (k)$, for several values of $m_X$ are shown in Fig.~\ref{fig:nk}
at the moment corresponding to 10 completed inflaton oscillations.
The particle momentum is measured in units of the inflaton mass.
In contrast to this, in the fixed de Sitter background
$4\pi^2\langle X^2 \rangle \approx H^2 \int d \ln k \,(k/H)^{3-2\nu}$ with
$(3-2\nu) \approx 2m_X^2/3H^2$ at small $m_X$, and consequently
$\langle X^2 \rangle \propto 1/m_X^2$.
Note that the power spectrum in the
fixed de Sitter background grows towards large values of $k$,
which is opposite to the behaviour of Fig.~\ref{fig:nk}.

Therefore, calculations which would be based on the customary 
procedure of matching a fixed
de Sitter background to a subsequent Friedmann stage would give wrong
results, with $\Omega_X \rightarrow \infty$ at
$m_X  \rightarrow 0$. 

Matching is also dangerous in the case of large $m_X$.
When the change is too abrupt, it generates artificial particles.
This may easily happen for $m_X > m_\phi$, see e.g. \cite{wrong_paper}
where excessive production was found.
At $m_X \agt m_\phi$ the number of created particles decreases exponentially
with $m_X$.

As Fig.~\ref{fig:nk} shows, the power spectrum of fluctuations
in X-particles is almost scale independent at small k if
$m_X/m_\phi \approx 1$. Therefore, if such particles constitute a considerable
fraction of dark matter, these fluctuations will be transformed into
isocurvature density perturbations at late times and can affect 
large scale structure formation. Isocurvature
fluctuations produce 6 times larger angular temperature fluctuations
in cosmic microwave background radiation (CMBR) for the same amplitude 
of long-wavelength density perturbations 
compared to the adiabatic case \cite{SS84}.
Therefore, to fit observations by a single spectrum,  
the mass fluctuation spectrum in isocurvature cold dark matter cosmology 
must be tilted (with respect to scale invariant spectrum) to favor
smaller scales.  

Let the power spectrum of the field
fluctuations be $k^3P_X(k) \propto k^\beta$. Fit to the second 
moments of the large-scale mass and cosmic microwave
background distributions requires $\beta \agt 0.25$. Models
with $\beta $ ranging from 0.3 to 0.6 were 
considered in  Refs. \cite{ML,Peebles}. 
It is interesting that 
the spectrum is indeed correctly tilted for $m_X$ which is somewhat 
larger than the inflaton mass, see Fig.~\ref{fig:nk}.
On the other hand we see that lighter particles, $m_X < m_\phi$,
with minimal coupling to gravity and $\Omega_X \approx 1$ 
are excluded.

\subsection{Preheating}

Considering gravitational creation of particles only, we would be left
today with an oscillating inflaton field dominating the 
Universe.\footnote{Exclusions from this rule exist if inflaton 
potential does not have
a minimum and energy density in the inflaton field after inflation 
decreases faster than in conventional matter \cite{NO}.}
To reheat the Universe we must couple the inflaton to some other fields.
In oscillating background and in flat space-time the number density of 
particles grows exponentially: parametric resonance is always effective,
see e.g. Ref. \cite{gmm}. This would lead to explosive decay of 
oscillations. However, in an expanding universe the resonance 
is blocked by the redshift,
which removes created particles out of the resonance bands.
Thus, coherent oscillations of e.g. axion \cite{axion-decay}, or 
moduli fields \cite{moduli-decay},  do not decay via parametric
resonance in the expanding Universe. It is interesting that the field
trapped in a (self-) gravitational well can burst in radiation, in principle
\cite{it87,GBK}, but limitations on the relation between density and 
size of the clump make it hard to achieve critical conditions when 
elaborated dark matter models, like the axion \cite{it87,KW} 
are considered (axion miniclusters are promising objects
in this respect though \cite{minicl}). 

Narrow width of the resonance band
is a prime reason of stability and such fields whill decay only when the
Hubble constant $H$ becomes smaller than the particle width $\Gamma$
(i.e. the life-time of individual particle becomes smaller than the age
of the Universe).
This results in reheating temperature 
$T_{\rm RH} \sim \sqrt{\Gamma M_{\rm Pl}}$. For a long time it was believed 
\cite{old_reheat} that this is the end of the story for the inflaton 
field, until it was realized  \cite{KLS94} that the resonance 
parameter $q$, Eq. (\ref{mchi}) can naturally 
be extremely large in the inflaton 
case\footnote{ For a discussion of the parametric
resonance induced by an inflaton field see also refs.~\cite{infl_res}.}. 
Indeed, the coupling constant is multiplied by the
ratio of initial amplitude of the field to its mass, squared, which
for the inflation is $\sim 10^{10}$. The resonance is broad and it is
impossible to red-shift the system out of the resonance.
Resulting explosive decay is not  actually a parametric resonance,
since resonance parameter can change by orders of magnitude just during 
one period of oscillation \cite{KT97a,KLS2}, but decay occurs
anyway because each period of oscillations the adiabaticity conditions are 
violated during some short time intervals \cite{KT97a,KLS2}.
It is possible
that the inflaton oscillations decay after only a dozen of oscillations.

Let us now consider this process and introduce a non-zero coupling 
of the inflaton and the $X$ fields.
For the sufficiently large value of $q$ in Eq. (\ref{mchi})
the gravitational production of particles is negligible. If the 
amount of produced particles is relatively small, 
the process of their creation still
can be considered as creation by external time varying background
(oscillating zero mode of the inflaton). However, production
can be very efficient at large $q$ and back reaction of produced particles
on the motion of the inflaton field has to be included. This restores
conservation laws and, as a matter of fact,
we are considering the decay of the inflaton into $X$ now. Here we restrict
ourselves to the case of scalar X-particles.

The equation of motion for the inflaton field is
\begin{equation}
{\varphi}'' - \nabla^2 \varphi - \frac{\ddot a}{a} \varphi 
+ a^2\varphi + 4q\chi^2\varphi = 0 \,\, .
\label{eqzm}
\end{equation}
Now the background (i.e. the inflaton field) is not homogeneous 
anymore and the complete 
quantum problem is not easy. However, the efficiency of particle creation
saves the day: the system rapidly becomes classical \cite{KT96}
and the problem can be handed over to a computer \cite{KT97a}-\cite{PR}.

Instead of the number density of created particles it is more convenient 
(and rigorous) to measure the related quantity: the
variance of the fluctuation field
\begin{equation}
\langle \chi^2 \rangle = \frac{1}{(2\pi)^3}
\int d^3 k |\chi_{\vec k}|^2 \,\, .
\label{nor}
\end{equation}

Initially fluctuations are small and the problem can be followed
in the Hartree approximation. Namely, the field $\chi$ in 
Eq. (\ref{eqzm}) is replaced by its average, $\langle \chi^2 \rangle$,
and spatial gradients of the inflaton field are neglected.
The formalism of section~\ref{quantum_fields} then applies. In the situation
of efficient particle creation
occupation numbers grow and at some point quantum averages
can be approximated by classical averages computed with the help of a certain
distribution function \cite{PS96,KT96}. At late times the problem becomes
classical and it has to be supplemented with appropriate initial conditions
which reflect early quantum evolution. 
Those are specified in the Fourier space as follows \cite{KT96,KT97a}.
Amplitudes of mode functions are distributed with the probability density
\begin{equation}
{ P}[\chi_{\vec k}] \propto \exp\left[ -\frac{2\phi^2(0)}{m^2} 
\omega_k(0) |\chi_{\vec k}(0)|^2 \right] \,\,  ,
\label{dis1}
\end{equation}
and mode functions have random phases.
The initial ``velocities" are locked to their corresponding ``coordinates"
\begin{equation}
{\dot \chi}_{\vec k} (0) = -i\omega_k(0)\chi_{\vec k} \,\, .
\label{vel}
\end{equation}
The parameter $m^2/\phi^2(0)$ sets the scale of $\langle \chi^2 \rangle$
which separates regions of quantum
and classical fluctuations in this model.
The semiclassical description is reliable as long as 
$g^2/4\pi=  q m^2/4\pi\phi^2(0)\ll 1$.

Full non-linear problem can now be solved numerically on the lattice 
(namely, Eq. (\ref{eqzm}) and corresponding equation for the field 
$\chi(\tau,{\bf x})$ are solved directly in the configuration space).
This classical problem accounts for all the effects of particle creation,
their rescattering, inverse decays, etc. Initial stage of intensive growth
of fluctuations (similar to parametric resonance, but not equivalent to it
since the resonance parameter $q$ changes rapidly in the expanding Universe) 
is followed by the stage which is similar to the 
Kolmogorov turbulence
when smooth power spectra of fluctuations are established which slowly
approach equilibrium \cite{KT96,KT97b,PR,Son}.

\begin{figure}
\centerline{\leavevmode\epsfysize=7.cm \epsfbox{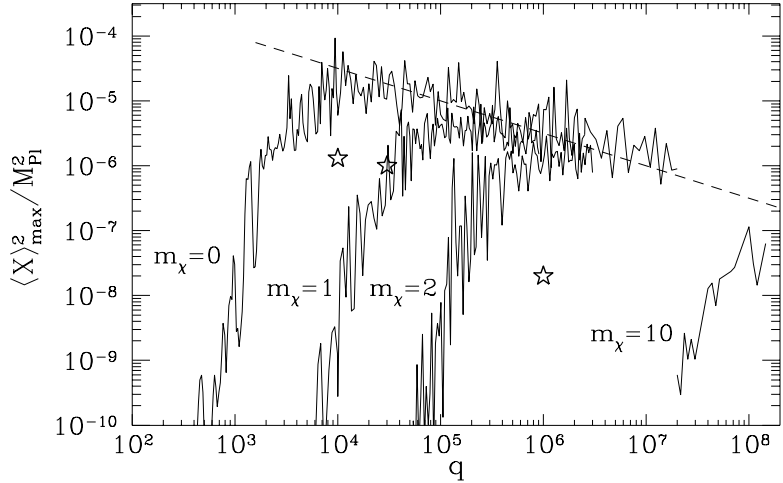}}
\vspace{10pt}
\caption{The maximum value of the variance $\langle X_{\rm max}^2 \rangle$ as 
a function of $q$. Solid lines correspond to the Hartree approximation with
different values of $m_\chi \equiv m_X/m_\phi$, Ref. \cite{KT97a}.
The stars are results of fully 
non-linear lattice calculations with $m_X \ll m_\phi$, Ref. \cite{KT97b}.}
\label{max_X}
\end{figure}

An important quantity is the maximum strength of fluctuations achieved during
the course of evolution (fluctuations are diluted by the expansion of the 
Universe after decay slows down). Compilation of results of Refs.
\cite{KT97a,KT97b} for the realistic case 
$m_\phi=10^{-6} M_{\rm Pl}$ is shown in Fig.~\ref{max_X} as a function of model
parameters. The stars are results of the lattice calculation which
takes all back reaction effects into account, solid lines correspond
to the computationally less expensive Hartree approximation.

To compare this with the gravitational creation note that in the Hartree
approximation $\rho_X/\rho \sim 2$ when fluctuations reach their maximum.
Even the lowest level of fluctuations shown in Fig.~\ref{max_X} exceeds
the amount of gravitationally created particles. However, while
stable particles will be generally overproduced, this mechanism still can be 
relevant for dark matter and UHECR phenomenon, when e.g. for a given value
of $m_X$ the value of $q$ is low enough to prevent effective
particle creation, but not negligible. Fig.~\ref{max_X}
shows that particles 10 times heavier than inflaton
are produced and even this is not a limit in the case of very large $q$
\cite{instant}.

In the opposite situation when $q$ is sufficiently large and fluctuations
develop to the full strength, an interesting and important 
phenomenon of non-thermal phase transitions \cite{nth} can occur 
which is the subject of discussion in the 
following section.

\section{Topological defects and inflation}

Decaying topological defect can naturally produce very energetic
particles, and this may be related to UHECR 
\cite{hsw,S,necl,vortons,monium,mon}, 
for recent reviews see \cite{bs}. However, among motivations for
inflation there was the necessity to get rid of unwanted topological defects.
And inflation is doing this job excellently. Since temperature after
reheating is constrained, especially severely in supergravity models, 
it might be that the Universe was never reheated up to the point
of  GUT phase transitions.
Topological defects with a sufficiently high scale of symmetry
breaking cannot be created.
Then, how could such topological defects populate the Universe?

The answer may be provided by non-thermal phase transitions \cite{nth}
which can occur in preheating after inflation. Explosive
particle production caused by stimulated decay of inflaton oscillations
lead to anomalously high field variances which restore symmetries
of the theory even if the actual reheating temperature is small. The defects
are formed when variances are reduced by the continuing  expansion of 
the Universe and a phase transition occurs.

\subsection{Non-thermal phase transitions}

Let us first describe ideas qualitatively.
Let the inflaton oscillations which have amplitude $\phi(0) \sim M_{\rm Pl}$
decay rapidly into some field $X$. In the instant process of decay 
the energy conserves
and we have $m_\phi^2  M_{\rm Pl}^2 \sim k^2 \langle X^2 \rangle$,
where on the left hand side of this equality we write the initial 
inflaton energy
and on the right hand side we write the final energy stored in the X-field.
The inflaton decays much faster than the system thermalizes, therefore
typical momentum of $X$ particles is of order of the inflaton mass, 
$k \sim m_\phi$.
We find $\langle X^2 \rangle \sim M_{\rm Pl}^2$. In thermal equilibrium
with the temperature $T$ we would have $\langle X^2 \rangle = T^2/12$.
In this respect the strength of fluctuations is the same as it would be
in equilibrium with the Plankian temperature despite the fact that the real
reheating temperature is much lower. Of course, this is an extreme estimate,
mainly because the expansion of the Universe was neglected.
The realistic $\langle X^2 \rangle$ is shown in Fig.~\ref{max_X}, but it 
can still exceed equilibrium temperature by many orders of magnitude.

This effect is important for the behaviour of spontaneously 
broken symmetries at preheating.
Let us consider the model in which in the vacuum state the symmetry 
is broken by an order parameter $\Phi$. At the tree level this can
be described by the potential 
\begin{equation}
V(\Phi ) = -\mu^2 \Phi^2/2 + \lambda_\Phi \Phi^4/4.
\label{symm_b_pot}
\end{equation}
The parameter $\mu$ is related to the symmetry breaking scale 
via $\mu^2 =\lambda_\Phi \Phi_0^2$. If the filed $\Phi$ and the product
of inflaton decay, $X$, are coupled (with corresponding interaction
term in the Lagrangian being $\alpha X^2\Phi^2/2$), 
at non-zero density of X-particles the effective mass of 
$\Phi$ field changes to
$m_{\rm eff}^2 = -\mu^2 + \alpha \langle X^2 \rangle$. 
The symmetry is restored if the effective mass became positive, 
$\langle X^2 \rangle > \lambda_\Phi \Phi_0^2/\alpha$.

The real problem is complicated and model dependent.
While some features can be anticipated
and some quantities roughly estimated, the issue requires numerical
studies. In recent papers \cite{1opt,strings,KK98} the defect 
formation and even
the possibility of the first order phase transitions during preheating
was demonstrated explicitly. 

\subsection{Topological defects in simple models}

Let us consider for simplicity the system when one and the same field
serves as the inflaton and the symmetry breaking parameter,
 $\Phi \equiv \phi$. We derive the set of models from the prototype 
potential
\begin{equation}
V(\phi ) = {\lambda \over 4}(\phi^{2}-v^{2})^{2} 
- {g^{2}\over 2} \phi^{2} X^{2}\, .
\label{td_pot}
\end{equation}
The inflaton scalar field $\phi$ has $M$ components,
$\phi^2=\sum_{i=1}^M \phi_i^2$, and 
interacts with an $N$-component scalar field $X$,~
$X^2=\sum_{i=1}^N X_i^2$. For simplicity,
the field $X$ is taken massless and without self-interaction.
The fields have minimal coupling to gravity in a FRW universe with a scale
factor $a(t)$. It is assumed that the inflaton oscillations start 
along $\phi_1$ direction.
In the effective mass of $\phi$ there are contributions
from $g^{2}\langle X^{2} \rangle$ as well as
$\propto \lambda\langle (\delta\phi)^{2}\rangle$.

We review several models, in the order of increasing complexity.
Initial conditions in all cases correspond to the vacuum for fluctuations.
System then evolved through particle creation, their rescattering, phase
transitions and finally defects creation.

\subsubsection{Domain walls {\rm \cite{wall}} }

\begin{figure}
\centerline{\leavevmode\epsfysize=7.cm \epsfbox{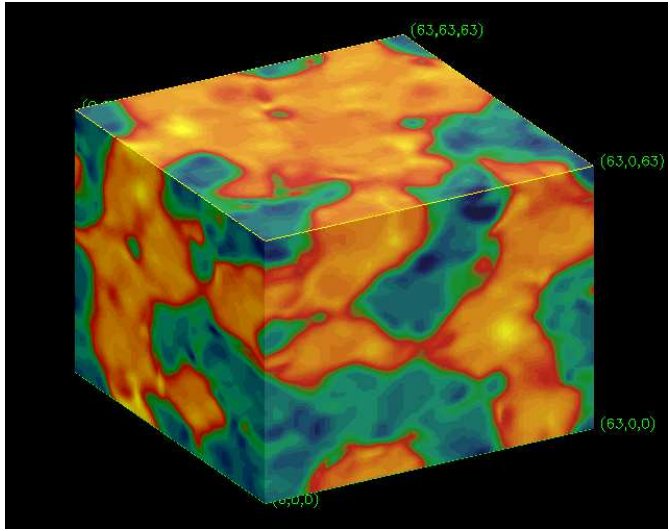}}
\vspace{10pt}
\caption{Domain wall structure generated in simulation of preheating
in the model Eq. (\ref{td_pot}) with $M=1$, $g=0$, Ref. \cite{wall}. }
\label{domains}
\end{figure}

Consider one component inflaton field, $M=1$, and no $X$-fields.
Relevant topological structures in the double-well potential
are domain walls.
Even in the model without fields $X$, fluctuations of the field
$\phi$  grow dramatically (because of the self-interaction 
$\propto \lambda$, the zero mode can decay in the process 
$2\phi \rightarrow 4\phi$; in addition to that the spinodal decomposition
is effective in the present model). This 
leads to formation of a domain structure at certain values
of $\lambda$. The domain structure which emerged in such situation is shown
in Fig.~\ref{domains}

\subsubsection{Strings {\rm \cite{strings} }}

Again, consider the model without $X$-fields, but the inflaton has two
components, $M=2$. Relevant topological structures in the Mexican hat 
potential are strings.

\begin{figure}
\centerline{\leavevmode\epsfysize=5.5cm \epsfbox{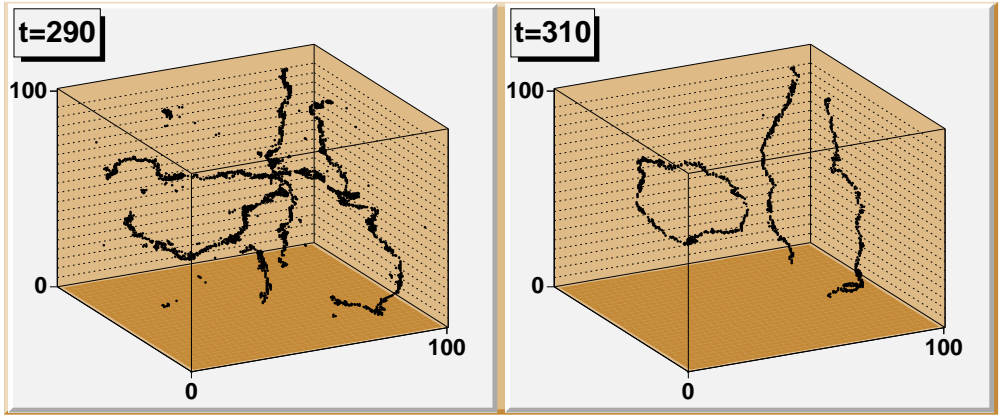}}
\vspace{10pt}
\caption{String network 
generated in simulation of preheating
in the model Eq. (\ref{td_pot}) with $M=2$, $g=0$ 
(see Ref. \cite{strings}) is shown
at two successive moments of time.}
\label{strings}
\end{figure}

Fluctuations of the fields $\phi_1$ and $\phi_2$ during preheating 
restore symmetry along $\phi_1$ direction, but along $\phi_2$
direction symmetry is broken. For some period of time the universe 
is divided into domains filled with the field $\phi_2 \approx \pm {\rm v}$. 
Gradually the amplitude of  fluctuations of the fields $\phi_i$
decreases, and symmetry $\phi_1 \to -\phi_1$ also breaks down.
At this moment the seed domain structure is transformed into a string network.
Fig.~\ref{strings} shows the string distribution in a simulation
with the symmetry breaking scale ${\rm v} =  3 \times 10^{16}$ GeV,
when a pair of ``infinite'' strings and one big loop were formed. 
Size of the box is comparable to the Hubble length at this time.

The final result is the string formation, but the sequence of symmetry breaking
patterns would be unusual for thermal phase transitions. For example,
the O(2) symmetry was only partially restored. Another peculiarity 
of non-thermal phase transitions is the 
possible presence of oscillating zero mode during final stages of the phase 
transition. This makes probability of formation of long strings 
(as well as domains in the previous model) to be a non-monotonic
function of $v$, at least for large $v$ and in simplest models considered
so far. In more complicated models inflaton oscillations can decay completely
before the phase transition, and the phase transition can be even
of the first order. Defect formation in such models will be more robust
and resembling more closely phase transitions in thermal background.

\subsubsection{First order phase transition {\rm \cite{1opt}} }

Here we consider one component inflaton field, $M=1$ and several $X$-fields. 
In the case $g^{2}/\lambda \gg 1$  the phase transition can be of the
first order. This can be expected by analogy with the usual
thermal case \cite{firstorder}.
The {\em necessary} conditions for this transition to occur and
to be of the first order are as follows.

(i) At the moment of phase transition a local minimum of the 
effective potential should be at $\phi =0$, which gives
$g^2 \langle X^2 \rangle > \lambda v^2$.

(ii) At the same time, the typical momentum $p_*$ of $X$ particles
should be smaller than $gv$. This is the condition of the existence of a
potential barrier. Particles with momenta $p<gv$ cannot penetrate the
state with $|\phi|\approx v$, so they cannot change the shape of the
effective potential at $|\phi| \approx v$. Therefore, if both conditions
(i) and (ii) are satisfied, the effective potential has a
local minimum at $\phi=0$ and two degenerate  minima at $\phi  
\approx \pm v$.

(iii) Before the minima at $\phi\approx \pm  v$ become deeper than the
minimum
at $\phi=0$, the inflaton's zero mode should decay significantly,
so that it performs small oscillations near $\phi=0$. Then, after
the minimum at $|\phi| \approx v$ becomes deeper than the
minimum at $\phi = 0$, fluctuations of $\phi$ drive the system over the
potential barrier, creating an expanding bubble.

All these conditions can be met more easily at large $g^{2}/\lambda \gg 1$
and with several $X$-fields. Results discussed
below were obtained \cite{1opt} on a $128^3$ lattice 
in the model with parameters $g^2/\lambda =200$ and
$v = 0.7 \times 10^{-3} M_{\rm Pl} \approx 0.8\times  10^{16}$ GeV, for a
two-component $X$, with the expansion  
of the universe assumed to be radiation dominated
(similar behaviour was observed in the
model with single $X$ field, i.e. $N=1$, as well).

\begin{figure}
\centerline{\leavevmode\epsfysize=7cm \epsfbox{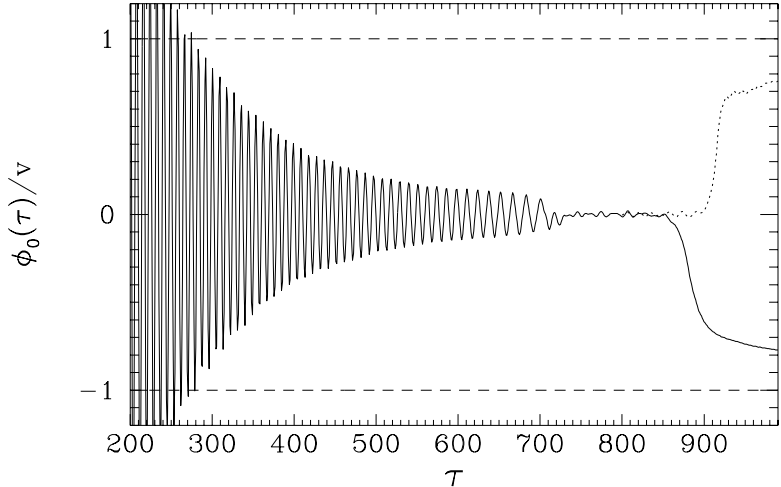}}
\caption{Time dependence of the zero-momentum mode of $\phi$ in
units of its vacuum value for two runs with different realizations
of random initial conditions for fluctuations in simulation of preheating
in the model Eq. (\ref{td_pot}) with $M=1$, $g \ne 0$, Ref. \cite{1opt}.
}
\label{fig:zmode}
\end{figure}

Time dependence of the zero mode $\phi_0$ is shown in Fig.~\ref{fig:zmode}.
Initially $\phi_{0}$ oscillates with a large amplitude ${\bar \phi}\gg v$.
If all fluctuations were absent, the zero mode
$\phi_0$, because of its dilution in the expanding universe, 
would at some time be unable to cross the potential barrier at 
$v=0$ and would start oscillations near
one of its vacuum values, $\pm v$. This would happen when the amplitude
of the oscillations becomes smaller than $\sqrt{2}v$. In Fig.  
\ref{fig:zmode} we see that
the actual dynamics is completely different. The zero mode of the
field  $\phi$ continues to oscillate near $\phi=0$ even when its
amplitude becomes much smaller than $v$.
In other words, the field oscillates on top of the local maximum of
the bare potential. This can occur only because the effective
potential acquires a minimum at $\phi = 0$ due to interaction
of the field $\phi$ with $\langle X^2\rangle$. And amplitude
of oscillations is decreasing much faster than it would decrease
due to the simple expansion of the Universe. This happens because 
of continuing decay of the zero mode into $X$-particles.

\begin{figure}
\centerline{\leavevmode\epsfysize=7.cm \epsfbox{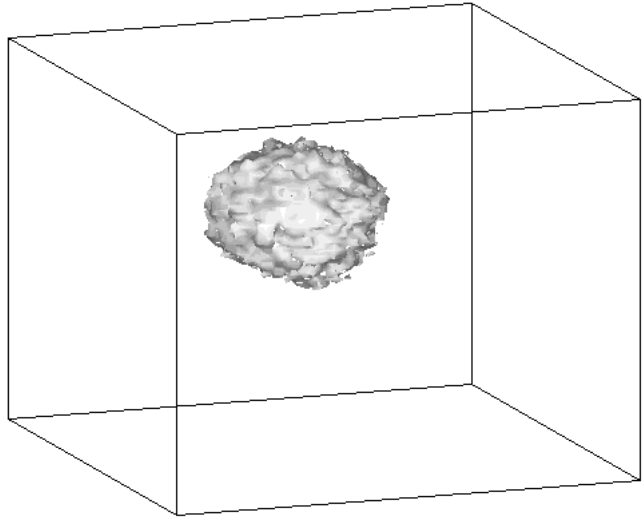}}
\vspace{10pt}
\caption{Spontaneously nucleated bubble of the new phase at the initial
stage of its expansion, Ref. \cite{1opt}.} 
\label{fig:ball}
\end{figure}

At still later times, $\tau \agt 720$ in Fig. \ref{fig:zmode}, the
zero mode of $\phi$ decays completely.   This  should  be  
interpreted as restoration of the symmetry $\phi \to -\phi$  by non-thermal
fluctuations. Finally, at $\tau \agt 860$ (when
$X$-fluctuations were diluted sufficiently by the expansion) 
a phase transition occurs
and the symmetry breaks down. In runs with different realizations for
initial (``vacuum'') fluctuations the system ends up either in $+v$
or in $-v$ vacuum and the transition happens at different time moments.
It was shown that transition is triggered in this model 
by a spontaneous  nucleation of a
bubble of the new phase and the
bubble's subsequent expansion  until it was occupying the whole
integration volume.
The field configuration at the beginning of the phase
transition is shown in Fig.~\ref{fig:ball}, where
the surface of the constant field
$\phi=-0.7 v$ is plotted at the beginning of the phase transition.
Inside the surface $\phi < -0.7 v$. (Minimum of the effective
potential is shifted somewhat from the vacuum value at this time,
see Fig.~\ref{fig:zmode},
because the contribution of fluctuations is still significant).

Models that exhibit behaviour shown in Fig. \ref{fig:zmode} will lead to 
the domain structure surviving until present. This would be a 
cosmological disaster and such a class of models is 
ruled out \cite{zko}.

However, different behaviour of the zero mode is observed at smaller values of
$g^2/\lambda$ or $N$ (but still $g^2/\lambda \gg 1$).
There, the phase transition can occur when the zero mode
still oscillates near $\phi=0$ with a relatively large amplitude.
This is a new, specifically nonthermal, type of a phase transition.
In such cases, bubbles of +$v$ and --$v$ phases will be nucleated in turn
(most often at the maximum of amplitude of the zero mode when it 
is in the closest proximity to the top of the potential barrier),
but their abundances need not be equal, and for certain values of the
parameters one of the phases may happen not to form  infinite  
domains. Such models are not ruled out and in fact may have 
interesting observable consequences, e.g. an enhanced (by bubble wall  
collisions) background of relic gravitational waves produced by the 
mechanism proposed in Ref. \cite{gw}. 

There are no reasons to doubt that in more complicated models non-thermal
phase transitions may lead to creation of magnetic monopoles, or
magnetic monopoles connected by strings (necklaces \cite{necl}), the latter
being the most favourable topological defect candidate for explanation of 
UHECR \cite{WB}.

\section{Conclusions}

Next generation cosmic ray experiments, like the Pierre Auger Project 
\cite{auger}, the High Resolution
Fly's Eye \cite{FlysEye}, the Japanese Telescope
Array Project \cite{JapArray}, and the Orbiting Wide-angle Light-collector 
(OWL) \cite{OWL}, will
tell us which model for UHECR may be correct and which has to be ruled out.

Very weakly interacting superheavy X-particles
with $m_X = ({\rm a~few}) \cdot 10^{13}$ GeV may naturally constitute a
considerable fraction of Cold Dark Matter. These particles are
produced in the early Universe from vacuum fluctuations
during or after inflation. 
Decays of X-particles may explain UHE cosmic rays phenomenon. 
Related density fluctuations 
may have left an imprint in fluctuations of cosmic microwave background 
radiation if scalar X-particles with minimal coupling to gravity
are approximately twice heavier than the 
inflaton and $\Omega_X \sim 1$.

If UHE cosmic rays
are indeed due to the decay of these superheavy particles,
there has to be a new sharp cut-off in
the cosmic ray spectrum at energies somewhat smaller than $m_X$. Since
the number
density $n_X$ depends exponentially upon $m_X/m_\phi$, the position
of this
cut-off is fixed  and can be predicted to be near
$m_\phi \approx 10^{13}$ GeV, the very shape of the cosmic
ray spectrum beyond the GZK cut-off being of quite generic form
following from the QCD quark/gluon fragmentation. 

Very discriminating signature is related to 
anisotropy of cosmic
rays. If particles immune to CMBR are there, the UHECR events
should point towards distant (i.e. beyond GZK sphere), 
extraordinary astrophysical sources \cite{FB98}.
If superheavy relic particles are in the game, 
the Galaxy halo will be reflected in anisotropy
of the UHECR flux \cite{DT98,BM98,AKENO}. If neither will be true but
UHECR will point instead to ``local'' galaxies with evidence for a
central supermassive black hole, that would imply that the existence
of a black hole dynamo is not a sufficient condition for the presence of
pronounced jets and UHECRs are created by the remnants of dead quasars
\cite{boldt}. If none of the above will be true, arrival directions will be
almost isotropic on large angular scales but UHECRs will cluster on small 
scales, then perhaps extragalactic magnetic fields are much stronger than 
previously thought \cite{LSB}, or perhaps cosmic necklaces do exist.

It is remarkable that we might be able to learn about the earliest 
stages of the Universe's evolution by studying UHECRs. 
Discovery of heavy relic
X-particles will mean that the model of inflation is likely correct, or that
at least standard early Friedmann evolution from the singularity is ruled out,
since otherwise X-particles would have been inevitably 
overproduced \cite{KT98,IT99}.

We conclude that the observations of Ultra High Energy cosmic rays
can probe
the spectrum of elementary particles in the superheavy range and can give
an unique opportunity for investigation of the earliest epoch of
evolution of the Universe,
starting with the amplification of vacuum fluctuations during inflation
through fine details of gravitational interaction and down to the physics
of reheating.

\section{Acknowledgements}

The work of V.K. was supported partially by  the 
Russian Fund for Fundamental Research grant 98-02-17493-a.

%\end{pf*}
\end{document}